\documentclass[preprint,12pt]{elsarticle}

\usepackage{amssymb}

\usepackage[nodots]{numcompress}

\usepackage{graphicx}
\usepackage{txfonts,enumerate}
\usepackage[hyphens]{url}

\journal{JQSRT}

\def\magn{mag$_{\rm SQM}$ arcsec$^{-2}$}

\def\mcdm{mcd m$^{-2}$}
\def\scotopiclimitmag{18.9 mag$_{\rm SQM}$ arcsec$^{-2}$}

\begin{document}

\begin{frontmatter}

\title{The night sky brightness at Potsdam-Babelsberg\\
including overcast and moonlit conditions (Note)}

\author[1]{Johannes Puschnig\corref{JP}}
\ead{johannes.puschnig@astro.su.se}
\author[2]{Axel Schwope}
\author[3]{Thomas Posch}
\author[2]{Robert Schwarz}

\cortext[JP]{Tel: +46 (0)8 5537-8533, Fax: +46 (0)8 5537-8510}

\address[1]{University of Stockholm, Universitetsv\"agen 10, SE-11418 Stockholm, Sweden}
\address[2]{Leibniz Institut f\"ur Astrophysik, An der Sternwarte 16, 
D-14482 Potsdam, Germany}
\address[3]{Universit\"at Wien, Institut f\"ur Astrophysik, 
T\"urkenschanzstra{\ss}e 17, A-1180 Wien, Austria}

\begin{abstract}
We analyze the results of 2 years (2011--2012) of night sky photometry performed at the Leibniz Institute for Astrophysics in Potsdam-Babelsberg.
This institute is located 23\,km to the southwest of the center of Berlin.
Our measurements have been performed with a Sky Quality Meter.
We find night sky brightness values ranging from 16.5 to 20.3 \magn;
the latter value corresponds to 4.8 times the natural zenithal night sky brightness.
We focus on the influence of clouds and of the moon on the
night sky brightness. It turns out that Potsdam-Babelsberg, despite its proximity to Berlin, still shows a significant correlation of the night sky brightness with the lunar phases. However, the light-pollution-enhancing effect of clouds dominates the night sky brightness by far: overcast nights 
(up to 16.5 \magn) are much brighter than clear full moon nights 
(18--18.5 \magn). 
\end{abstract}


\end{frontmatter}


\section{Motivation \label{sec:intro}}

Astronomers have been monitoring the brightness of the night sky for many decades at several observatories around the world (see, e.g., 
\citet{Patat2003}, \citet{Patat2006} and \citet{Patat2008}).
When it comes to choosing the site of a new ground-based telescope, measurements of the night sky brightness are of crucial importance in order to find sufficiently dark sites with as many clear nights per year as possible. Much less investigations have been devoted, for obvious reasons, 
to the sky brightness at abandoned astronomical sites
and in urban areas both under clear and overcast conditions.

Recently, however, monitoring programmes have been initiated in many places to study the brightness of the clear {\em and}\/ cloudy night sky as well as during moonless {\em and}\/ moonlit nights. This is because of the increasing
knowledge about the influence of light at night not only on astronomical observatories, but also on ecosystems, on chronobiological rhythms and on our society. In October 2013, the first international conference on artificial light at night -- covering all these topics -- took place (\citet{Krop-Benesch2013}).
The present paper is indeed motivated by the fact that light pollution needs to be monitored continuously due to its implications for astronomy, energy 
consumption, wildlife and human health (e.g., \citet{Kyba2012a},
\citet{Davies2013}, and references therein).

Our study of the night sky brightness (henceforth: NSB) at Potsdam-Babelsberg presents a special case in this context. Potsdam-Babelsberg is located 3\,km away from the center of Potsdam (population: 159 500) and 23\,km from the center of 
Berlin (3,375 million inhabitants). During clear nights, the NSB measured at
Potsdam-Babelsberg is mainly influenced by the proximity of Berlin with its
huge light dome. For those numerous nights, however, in which the sky is covered by clouds -- especially clouds lower than cirrus -- the sum of the upward directed artificial light from Potsdam is most probably the largest source of the night sky brightness that we measure. A third case is of interest as well: clear nights close to full moon. As \citet{Davies2013} did for the first time for Plymouth (UK) and its surroundings, we examined whether or not the dependence of the mean NSB on the phase of the moon is still significant at the observatory of Potsdam-Babelsberg.


\section{Measurement site and method \label{sec:methods}}

All our NSB measurements were performed on top of the ``Schwarzschildhaus'', a building that is located 160\,m to the west of the Potsdam-Babelsberg observatory. The location of our measurement device is sufficiently high to preclude any {\em direct}\/ irradiation of artificial light on our detector. Hence, it is indeed only the light scattered by the night sky which we measure.
The question of single versus multiple scattering of sunlight during the
twilight and the transition from dominant scattering of sunlight to the dominant scattering of artificial light will shortly be addressed in Sect.\ 3.

The geographical position of our measurement site is summarized in Tab.\ 1.

\begin{table}[h]
\caption{Geographical coordinates of our measurement site Potsdam-Babelsberg.}
\begin{center}
\begin{tabular}{| c | c | c |}
  \hline
	Altitude & Longitude               & Latitude                \\ \hline
	70m      & 13$^{\circ}$ 06' 06' E  & 52$^{\circ}$ 24' 18'' N \\
  \hline
\end{tabular}
\end{center}
\label{t:sites}
\end{table}

NSB measurements at Potsdam-Babelsberg were initiated in October 2010.
However, we decided to use only the data from the completed years 2011 
and 2012 for our analysis.

We used Unihedron's Sky Quality Meter with an integrated lensing system. The model we used is called ``SQM-LE'', but for the sake of brevity, we will refer to it as ``SQM'' here. The instrument was placed into a weatherproof housing and directed into the zenith, thus detecting the zenithal and near-zenithal sky brightness. The device is connected to a computer via ethernet. The 
sampling rate we used was 2.1\,sec.

The spectral sensitivity of the SQM, which has been studied by \citet{Cinzano2005}), differs from the transmittance of a Johnson V filter in three respects:

\begin{itemize}

\item{The peak of the SQM sensitivity is at 540\,nm.}

\item{While the Johnson V filter largely cuts off radiation with a wavelength
smaller than 470\,nm, the SQM detects light down to wavelengths smaller 
than 400\,nm.}

\item{Beyond 550\,nm, the SQM sensitivity and the Johnson V transmittance are fairly similar, but the SQM has a broader transmittance window also at these wavelengths.}

\end{itemize}

These differences motivate us to use the unit ``\magn'', which has also been proposed by \citet{Biggs2012}. The difference between \magn\ and  mag$_{\rm V}$ arcsec$^{-2}$ is colour-dependent and lies between 0.2 and 0.5\,mag according to \citet{Cinzano2005}.

However, a more systematic investigation of the influence of detector
temperature, detector aging, zero-point-differences between individual
SQMs etc. still needs to be done in future.


\subsection{Data storage and web interface}

All data presented and analyzed in this paper can be found as plots on the website \url{http://verlustdernacht.aip.de}. The plots of the zenithal NSB shown on this website will be called 'scotograms' in  the following, derived from the Greek word 'skotos' = dark.


\section{Data analysis}


\subsection{Comparison to literature values for the twilight sky brightness}

As a 'reality check' for our data, we compared them both to a table
of twilight sky brightness values published in the 1965 edition of the 
'Landolt-B\"ornstein' (LB) encyclopaedia and to the single-scattering
model of the twilight sky brightness published by \citet{Patat2006}.
Both the LB data set (\citet{Hellwege1965}) and the paper by Patat et al.\ 
refer to the brightness of the twilight and night sky without light pollution.
Another goal of our comparison was thus to find out for which depth of the Sun below the horizon the anthropogenic contribution to the measured NSB becomes noticeable or even dominant.

We converted the LB zenithal luminance data into zenithal sky brightness data 
in mag arcsec$^{-2}$ by using the equation:

\begin{equation}
{NSB}\,[mag_{V} arcsec^{-2}] = 
12.6 - 2.5 {\rm log}\,({luminance}\,[cd/m^2])
\label{eq:unihedron}
\end{equation}

We refer to (\citet{Cinzano1997}, p.\ 112) and (\citet{Garstang1986}, eq.\ 19) for the derivation. Note that this relation has been derived for the NSB measured in {\rm mag$_{V}$ arcsec$^{-2}$}, while it is only an approximation 
for the NSB measured in \magn.

\begin{figure}
	\centering
	\includegraphics[width=11.5cm]{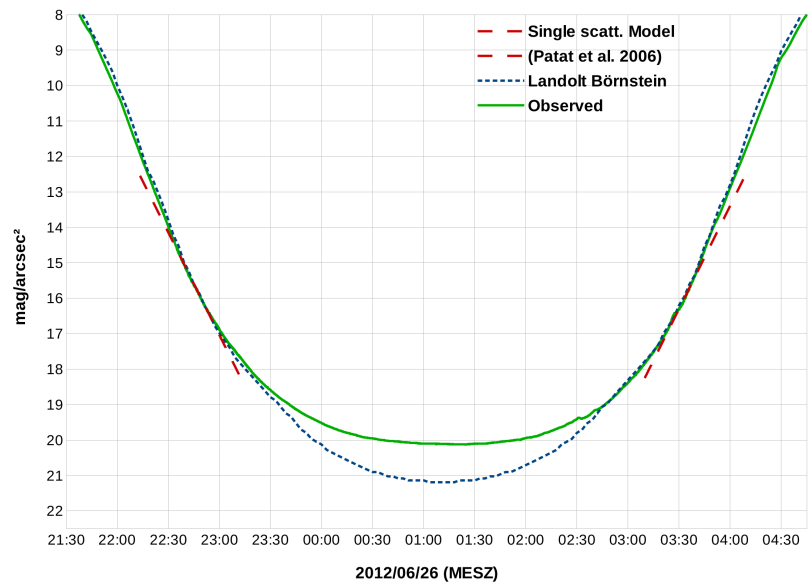}
	\caption{Comparison between standard literature values of the zenithal NSB
	(after Landolt-B\"ornstein, dotted line) and the NSB actually measured at 
	Potsdam-Babelsberg during the clear night from 26th to 27th of June 2012
	(solid line). The linear approximation to the single-scattering  
	model of the twilight sky brightness used by \citet{Patat2006} is also 
	overplotted for comparison, but only for solar elevations of 
	--5$^{\circ}$ to --10$^{\circ}$ (dashed line).
	}
	\label{fig:comparison}
\end{figure}

Figure \ref{fig:comparison} shows a scotogram measured at Potsdam-Babelsberg during a very clear night close to the summer solstice, namely from 26th to 27th June 2012, compared to the NSB which was expected based on the LB data. It can be seen that for an NSB larger (darker) than 18\,\magn, the measured values begin to deviate from the LB values due to the influence of light pollution.
This zenithal sky brightness is reached when the Sun reaches a depth of 10.2$^{\circ}$ below the horizon. The darkest NSB recorded during this night was 
20.1\,\magn. This in turn corresponds to the natural NSB predicted for 
h$_{sun, equiv}$ = --12.7$^{\circ}$ in the case of a light-pollution-free sky, while the real value was h$_{sun, real}$ = --14.3$^{\circ}$ at midnight at our location.

Given that an NSB of 20.1\,\magn\ is rarely being surpassed at Pots\-dam-Babelsberg even during clear, moonless nights,
our results can also be expressed in the following way: {\em The skyglow
at Potsdam-Babelsberg corresponds to permanent nautical twilight. The
sky luminance range of astronomical twilight is barely reached 
at this location.}

In Fig.\ \ref{fig:comparison}, we also compare our selected clear-night scotogram to a linear approximation to the single-scattering model used by \citet{Patat2006}. Patat et al.\ have pointed out that the brightness of the clear twilight sky in the V band is dominated by contributions of single scattering until h$_{sun}$ = --5$^{\circ}$. For larger depths of the Sun,
the contribution of multiple scattering becomes much more significant, which is reflected in Fig.\ \ref{fig:comparison} by the deviation of the measured and LB curves from the {\em lower} end of the dashed lines towards higher NSB values. Only for h$_{sun}$ = --15$^{\circ}$ and below, as pointed out again by Patat 
et al., the emission components (airglow etc.) instead of the scattering components become the dominant contributions to the NSB. But this phase cannot be seen in Fig.\ \ref{fig:comparison}, even for two reasons: first because 
h$_{sun}$ = --15$^{\circ}$ is not reached during this summer night and 
second because the anthropogenic contribution to the NSB outshines the 
airglow component during {\em every}\/ night at Potsdam-Babelsberg 
(see the difference between the solid and the dotted line).

Nevertheless, our results indicate that Potsdam-Babelsberg provides still {\em relatively}\/ dark skies relative to the conditions prevailing in the suburbs of other industrialized cities with several million inhabitants. This can be concluded, e.g., by a comparison of our results with those derived by \citet{Biggs2012} for Perth.


\subsection{Typical ranges of the night sky brightness at Babelsberg}

The next question of interest is: if we include overcast nights and nights with bright moonlight, how bright can nights get at Potsdam-Babelsberg? What is the full range of {\em mean} NSBs?

We calculated the mean NSBs for each night in the following way:

\begin{equation}
{<NSB>}\,[mag_{SQM} arcsec^{-2}] =
\frac{\sum \limits_{t=1}^n a_{t}}{n},
\label{eq:meanmag}
\end{equation}

using only measurements $a_{t}$ within the time interval [1,n] starting from the end of the nautical twilight when the sun's height was lower than 
--12$^{\circ}$ until the beginning of the nautical twilight at time step n.
By a statistical analysis of more than 14 million individual measurements from 2011 and 2012, we found the following results:

\begin{itemize}

\item{For clear and moonless skies, the mean night sky brightness at 
Potsdam-Babelsberg can reach values down to 20.3 \magn. Referring to a natural 
{\em zenithal}\/ night sky brightness of 22.0 mag arcsec$^{-2}$
(lower limit to the data measured by \citet{Patat2006}),
20.3 \magn\ corresponds to 4.8 times the natural sky brightness.}

\item{For clear nights close to full moon, we typically find mean NSBs 
of 18.5 \magn, corresponding to 4.3 mcd/m$^2$, with maximum NSB values of up to 18.0 {\magn} close to the full moon's culmination. Cloudy nights close to full moon can be still brighter than that due to enhanced scattering.}

\item{For clouded to overcast skies, mean NSB values up to 16.5 \magn\ can 
occur due to the enhanced backscattering of urban lights. 
Typical values of the NSB for nights with clouds over Potsdam-Babelsberg 
are around 17.5 \magn.}

\end{itemize}

The most instructive way to show NSB ranges which we found so far are density plots. We did such plots for 2011 and 2012 (Figs.\ 2--3). Each of these figures contains about seven million individual data points.

\begin{figure}
        \centering
        \includegraphics[width=11.0cm]{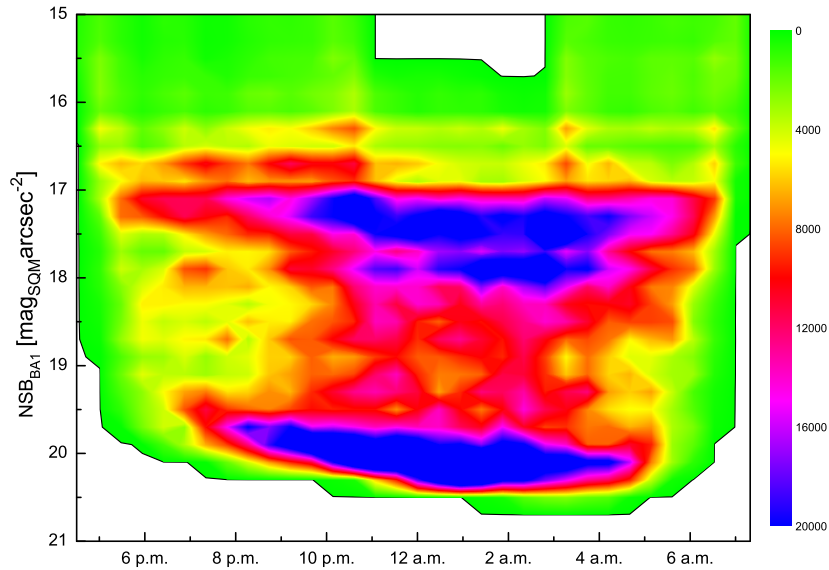}
        \caption[Density plot of the SQM measurements 2011]{Density plot of the SQM measurements performed at Potsdam-Babelsberg in 2011. Two dominant NSB ranges can be seen, one for cloudy and 
	overcast skies (16.5--18 \magn) and one for clear skies and small influence of the moon (19.5--20.3 \magn). A blurring of 0.2 mag was applied
	to the plot to eliminate quantization artifacts induced by the SQM device itself, according to indications of the manufacturer\footnotemark.}
        \label{fig:BA-darkness1}
\end{figure}

\begin{figure}
        \centering
        \includegraphics[width=11.0cm]{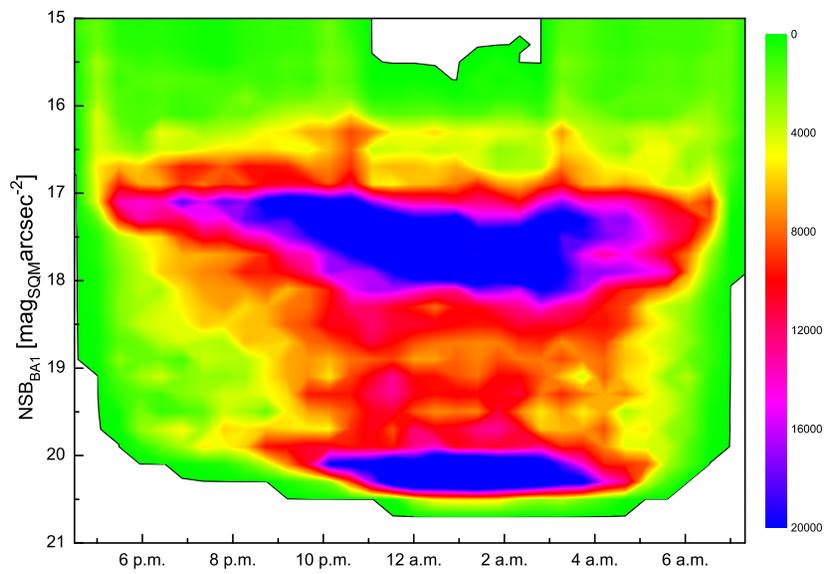}
        \caption[Density plot of the SQM measurements 2012]{Density plot of the SQM measurements done at Potsdam-Babelsberg in 2012. All times are local time, as in the previous figure.
	A blurring of 0.2 mag was applied to the plot to eliminate quantization artifacts induced by the SQM device itself, according to indications
	of the manufacturer.}
        \label{fig:BA-darkness2}
\end{figure}
\footnotetext{Only devices with firmware earlier than 4.3.21 are affected.}

The most conspicuous feature of Figs.\ \ref{fig:BA-darkness1}-\ref{fig:BA-darkness2} is the bifurcation between clear and cloudy nights. Clear nights turn out to be darker by almost 3 magnitudes -- a factor of 15.85 in luminance! -- than overcast nights. Without artificial illumination, overcast nights would be {\em darker}\/ than clear nights.

The decrease of the NSB towards later hours is less evident in the Potsdam-Babelsberg data than in the data from the Vienna University Observatory. While we found a typical gradient of 0.1 \magn\ per dark hour for the latter site, the present data sets indicate 0.08 \magn\ per hour for clear nights at Babelsberg, and still less for overcast nights.


\subsection{Variation of the night sky brightness with the lunar cycle}

Due to the publications of \citet{Puschnig2013} and \citet{Davies2013}, the following question came more into the focus of research: to which extent is 
the natural variation of the night sky brightness with the lunar phase still significant in urban congestion areas? As mentioned in Sect.\ 1, this question may have far-reaching consequences for wildlife and even for human health, since many species are known to synchronize a part of their behaviour with the full moon (see, e.g., \citet{Tessmar-Raible2011}; \citet{Zantke2013}; \citet{Longcore2004}).

\citet{Puschnig2013} have shown that the NSB variation with the lunar phases
is close to extinction at a distance of 3\,km from the center of Vienna (1.8 million inhabitants). \citet{Davies2013} have shown that the natural rhythm of the lunar sky brightness is no longer evident in the center of Plymouth 
(258.000 inhabitants).

\begin{figure}
        \centering
        \includegraphics[width=9cm]{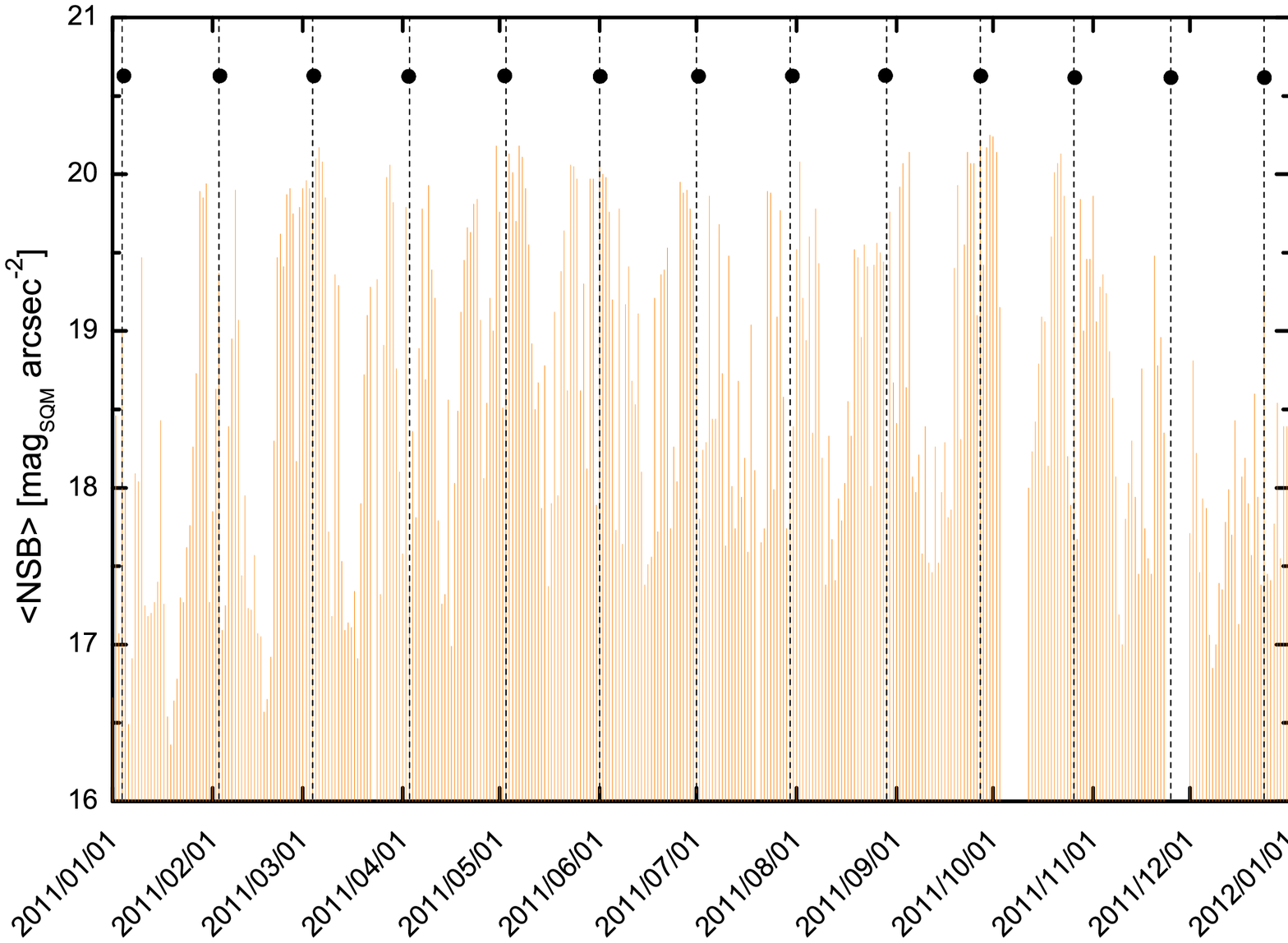}
	\includegraphics[width=9cm]{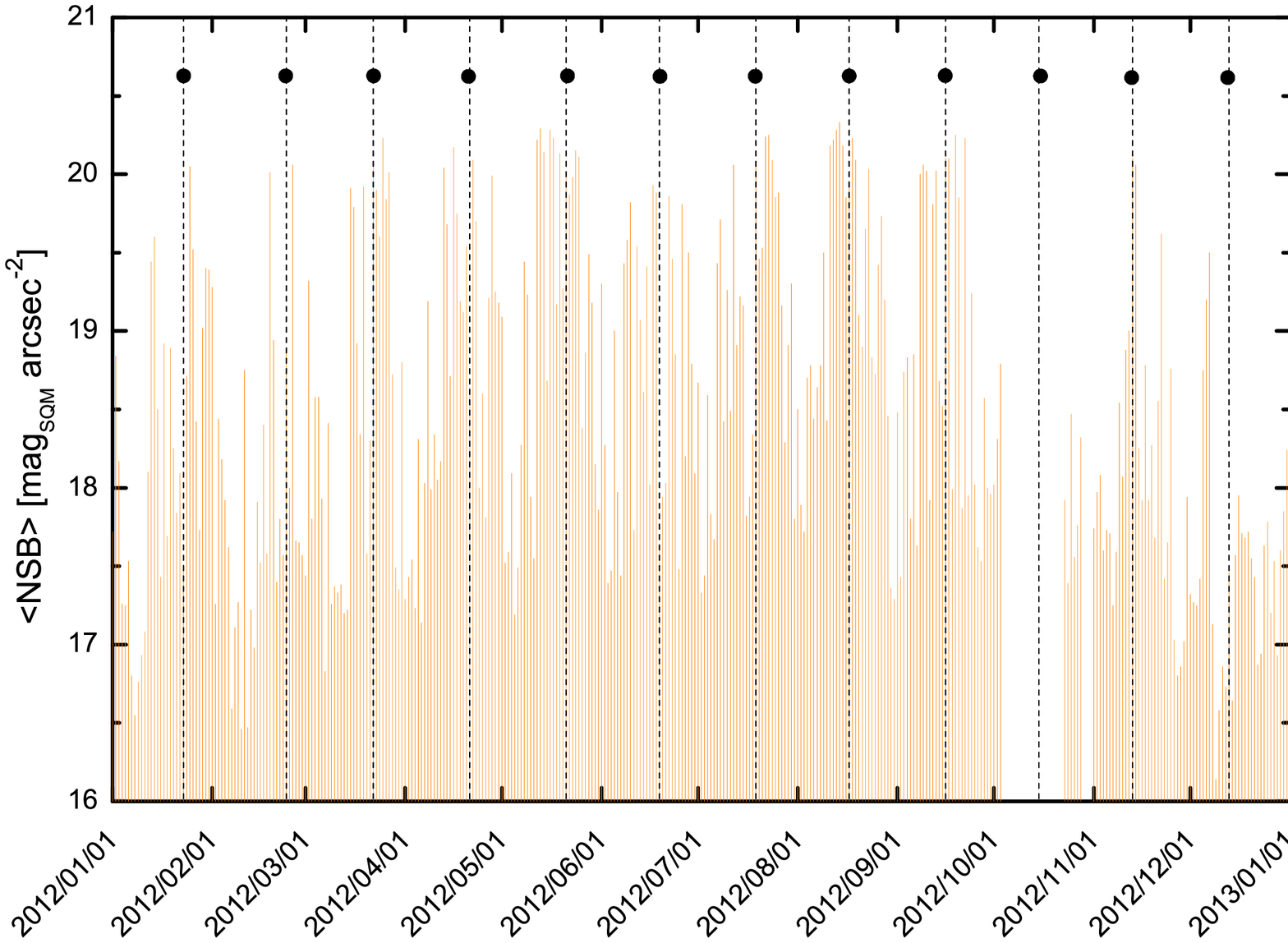}
        \caption{Mean NSBs for all nights in 2011 and 2012 for which our SQM at
	Potsdam-Babelsberg delivered data. A moon phase periodicity is still clearly
	seen in both plots, as the black filled circles and dashed vertical lines 
	(= times of new moon) indicate.}
        \label{fig:meanval}
\end{figure}

For Potsdam-Babelsberg, the calculation of mean NSB values for each night in 2011 and 2012, except for some nights where our device did not deliver any data, yields the pattern shown in Fig.\ \ref{fig:meanval}. The brighter full moon nights can clearly be identified in both plots. However, one can also see that there are many nights close to new moon where the mean NSB reaches values close to those typical for full moon.

In summary, we do detect a periodicity of the NSB dependent on the lunar phase, but again a highly weakened one due to the influence of artificial light.

\subsection{Variations of the sky brightness within a night}

\begin{figure}[h]
        \centering
        \includegraphics[width=11cm]{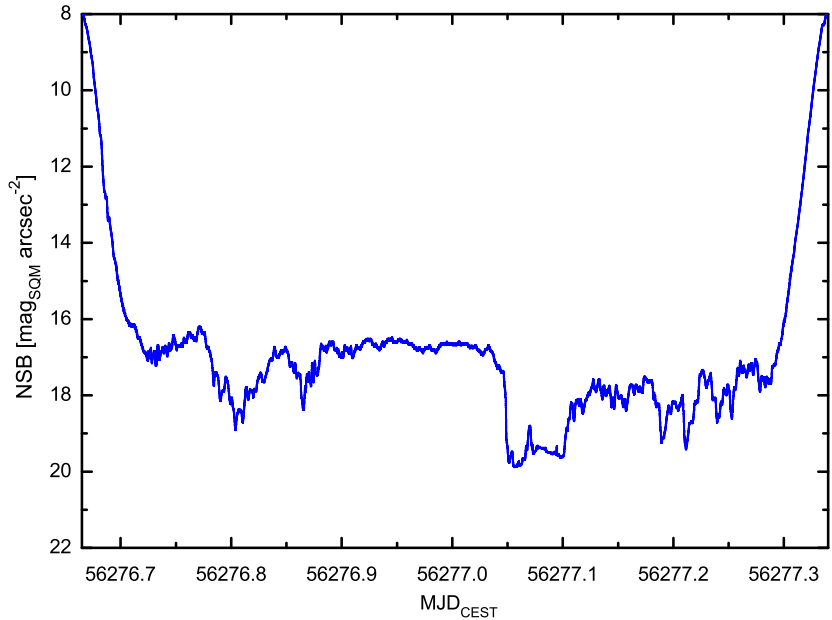}
        \caption{Scotogram obtained during the night from Dec 15 to Dec 16, 2012 showing variations of up to 3 \magn\ due to overcast sky.}
        \label{fig:nightvar}
\end{figure}

During cloudy nights, the variation of the NSB typically amounts to 1 
magnitude, but with strongly varying cloud coverage (see Fig.\ \ref{fig:nightvar}), it may also reach 3 magnitudes, corresponding to a factor of more than 
16 in the sky brightness. As mentioned in Sect.\ 1, for overcast skies, most of the contribution to the measured NSB is expected to be scattered upward light from Potsdam, while for overcast skies, the NSB is much more strongly governed by the light dome of Berlin. Figure \ref{fig:nightvar} thus shows a multiple transition between these two limiting cases within one night.

\subsection{Fraction of nights darker than 3 \mcdm at the zenith}

\begin{figure}
	\centering
	\includegraphics[width=11.5cm]{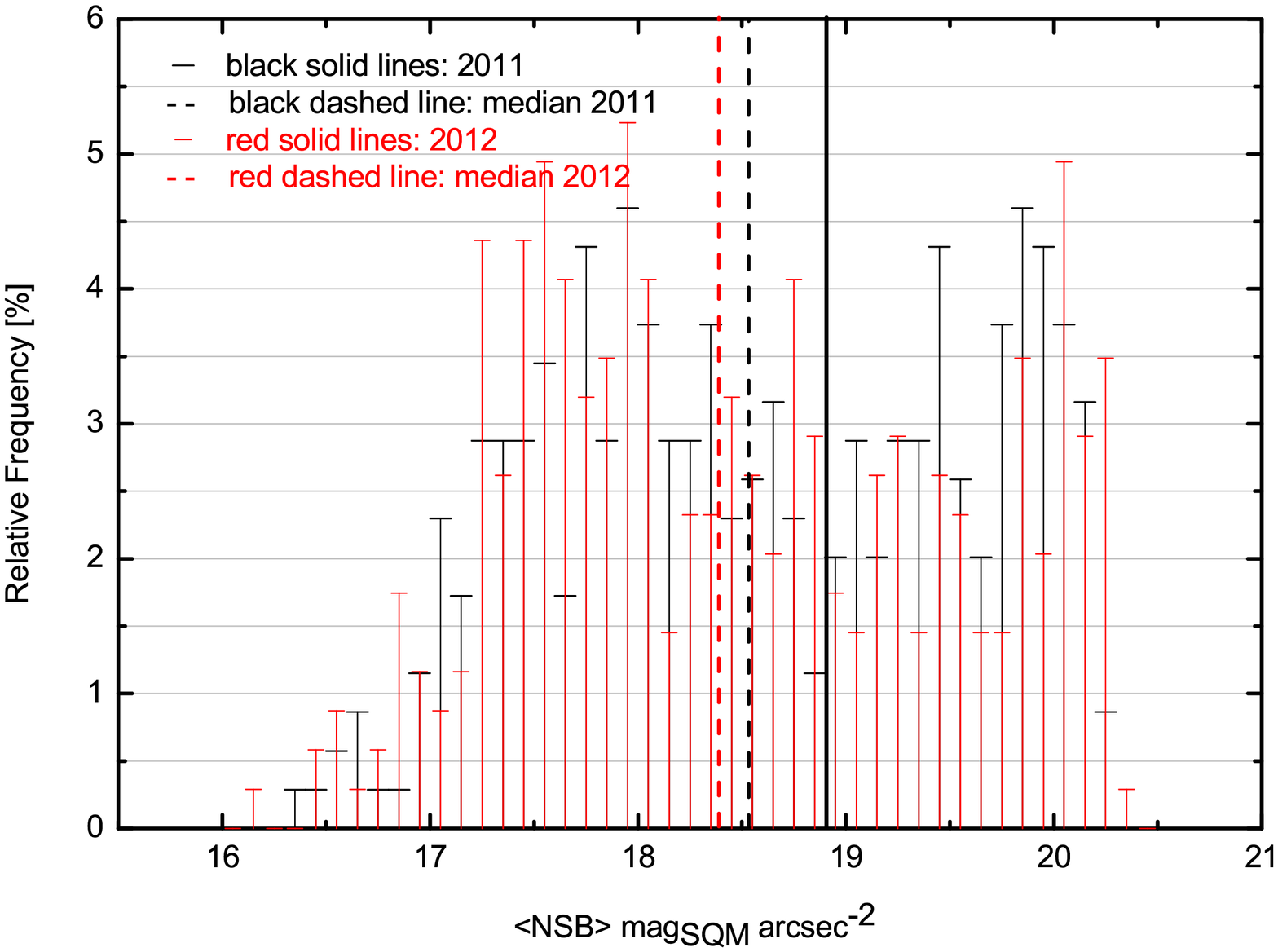}
	\includegraphics[width=11.5cm]{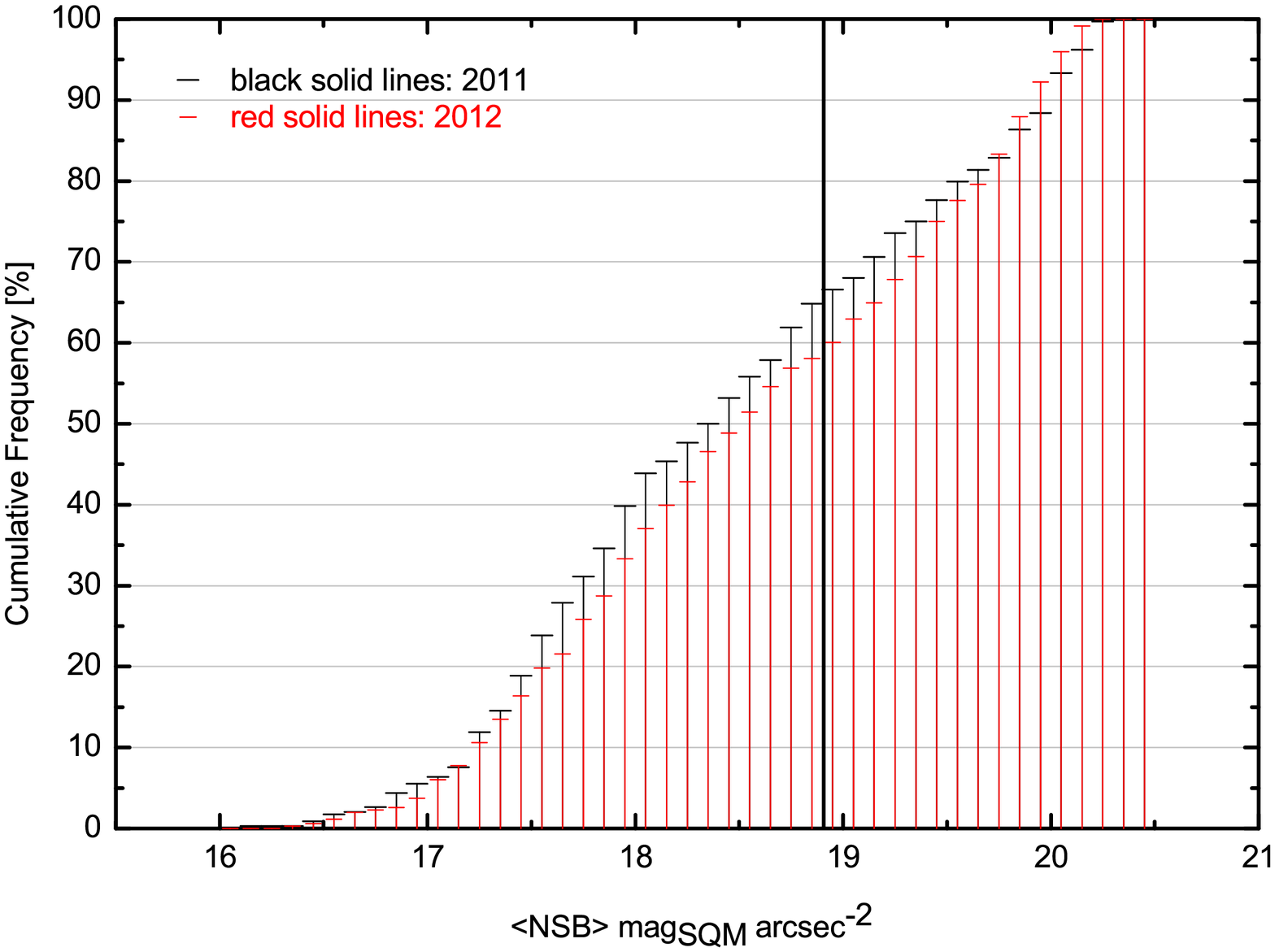}
	\caption{Histogram showing the relative (top) and cumulative (bottom) distribution of the mean NSB that we measured in 2011--2012. The vertical line corresponds to \scotopiclimitmag.}
	\label{fig:histo2011}
\end{figure}

\citet{Puschnig2013} have examined the fraction of nights in Vienna during which the average sky brightness is beyond 18.9 \magn, corresponding to a zenithal luminance of 3 \mcdm and found a value of only 10\%.
At Potsdam-Babelsberg, 40\% of all nights have mean NSB values 
$>$18.9 \magn. This number includes nights with cirrus and other thin clouds, 
i.e.\ it is not the percentage of astronomically useful nights. Nevertheless, this is a quite high fraction of relatively dark nights for a site close to
a city with more than 3 million inhabitants. This result is likely to be related to the fact that the Berlin-Potsdam area is more moderately lit
(with a lumen output close to 700\,lm per capita) than other European metropolitan areas of comparable size.
Figure \ref{fig:histo2011} shows the distribution of our derived mean 
NSB values in the form of histograms.

\section{Conclusions}

Based on photometric measurements performed 2011-12 with a Sky Quality Meter at Potsdam-Babelsberg, we find that the most decisive factor for the night sky brightness is -- as for most urban and suburban sites nowadays -- the cloudiness of the night sky and not the phase of the moon.
The variation of the night sky brightness with the lunar phases, however, is still seen at Potsdam-Babelsberg, even though in a much less pronounced form than at sites with dark skies.

Overcast skies with their enhanced backscattering of city lights lead to a mean NSB of up to 16.5 \magn\ at Potsdam-Babelsberg, corresponding to a zenithal luminance of 27.1 \mcdm\ or 157 times the natural zenithal sky brightness.
According to \citet{Garstang2007}, nocturnal clouds over city centers can
reach luminances of up to 1800 \mcdm\ under certain conditions: high ground reflectivity, high optical depth of the clouds, light output of 1200\,lm per inhabitant.

Under clear and moonless skies and during the late night hours, the mean NSB may reach 20.3 \magn\, corresponding to 0.82 \mcdm\ or 4.8 times the natural zenithal sky brightness. In contrast, clear full moon nights at our site have a typical zenithal luminance of 18.5 \magn\ or 4.3 \mcdm\ or 25 times
the natural zenithal sky brightness.

Especially for clear and moonless skies, we detected a continuous decrease in the NSB -- without any major ($>$ 0.1 mag) regularly occurring discontinuous 
steps at any given time (such steps would be expected for a partial switch-off of artificial lights, e.g., around midnight).
The decrease in the NSB, or improvement of the sky quality, is found to be no larger than 0.08 \magn\ per hour. Only a very small, if any, contribution to this decrease may be from a natural darkening of the night sky in the V band after the end of the astronomical twilight (see \citet{Patat2003}).

Compared to the results presented by \citet{Puschnig2013}
for the Vienna University Observatory, the average skies over Potsdam-Babelsberg are significantly darker, namely by more than 1 magnitude, which is explainable by the larger distance of Potsdam-Babelsberg from the respective metropolis' center. At the same time, the {\em difference}\/ between the brightest and the darkest mean NSB values is smaller for Potsdam-Babelsberg (3.8 \magn) than for the Vienna University Observatory (4.25 \magn). We expect that {\em sites with larger light pollution tend to have larger spreads between the respective darkest and brightest NSB values}\/ than sites with lower levels of light pollution. Table \ref{t:summary} summarizes a part of our conclusions.

\begin{table}[h]
\caption{Some benchmark values of the night sky brightness and luminance
which we derived for Potsdam-Babelsberg. The assumed value of the natural zenithal NSB is to be understood as a ``darkest possible'' reference value, measured only at excellent sites and during minima of the solar cycle (see \citet{Garstang1986}, \citet{Patat2003}, \citet{Patat2003} and 
\citet{Patat2008}.}
\begin{center}
\begin{tabular}{| l | r| r |}
	\hline
	Sky condition & Typical mean & Corresponding \\
        & NSB [mag$_{SQM}$/arcsec$^2$] & luminance [\mcdm] \\ 
	\hline
	Moonless clear skies & 19.5--20.3 & 0.82--1.73 \\
	Clear skies with full moon & 18.0--18.5 & 4.3--6.9 \\
	Overcast skies & 16.5--18.0 & $\le$27.1 \\
	Natural zenithal NSB & 22.0 & 0.172  \\
	\hline
\end{tabular}
\end{center}
\label{t:summary}
\end{table}


\vspace{1.0cm}

\begin{small}

{\noindent \bf Acknowledgements}

\noindent 
This work was supported by the project ``Verlust der Nacht'' 
(funded by the Federal Ministry of Education and Research, Germany, 
BMBF-033L038A) as well as by the EU COST project ES 1204
``Loss of the Night Network''. JP acknowledges financial support from
the provincial government of Upper Austria.
The authors wish to thank Anthony Tekatch (Unihedron) for fruitful discussions and two anonymous referees whose comments helped to improve the original version of the paper.

\end{small}


\bibliographystyle{jqsrt}
\bibliography{bibliography}


\end{document}